\documentclass[a4paper,10pt]{article}

\usepackage{INTERSPEECH2020}
\usepackage{hyperref}

\title{End-to-end speech-to-dialog-act recognition}
\name{Viet-Trung Dang, Tianyu Zhao, Sei Ueno, Hirofumi Inaguma, Tatsuya Kawahara}
\address{School of Informatics, Kyoto University, Kyoto, Japan}
\email{trungdv@sap.ist.i.kyoto-u.ac.jp}

\begin{document}

\maketitle

\begin{abstract}
  Spoken language understanding, which extracts intents and/or semantic concepts in utterances, is conventionally formulated as a post-processing of automatic speech recognition. It is usually trained with oracle transcripts, but needs to deal with errors by ASR. Moreover, there are acoustic features which are related with intents but not represented with the transcripts. In this paper, we present an end-to-end model that directly converts speech into dialog acts without the deterministic transcription process. In the proposed model, the dialog act recognition network is conjunct with an acoustic-to-word ASR model at its latent layer before the softmax layer, which provides a distributed representation of word-level ASR decoding information. Then, the entire network is fine-tuned in an end-to-end manner. This allows for stable training as well as robustness against ASR errors. The model is further extended to conduct DA segmentation jointly. Evaluations with the Switchboard corpus demonstrate that the proposed method significantly improves dialog act recognition accuracy from the conventional pipeline framework.
  \end{abstract}
  %
  %
  \section{Introduction}
  \label{sec:intro}
  
  Spoken language understanding (SLU) has an important role in designing intelligent machines or robots that can take part in conversation with human. It is traditionally designed as a pipeline of an automatic speech recognition (ASR) component to transcribe speech into text and a natural language processing (NLP) component to extract meaning from the transcribed text. Each model is designed and trained independently. There are many specific domains such as weather information and train information, in which intents and slots are defined in a domain-dependent manner. In these tasks, the intent of an utterance from a user is first identified and then necessary slots are filled. This scheme has been widely deployed in many smart systems with voice user interface such as smartphone assistants and smart speakers.
  
  On the other hand, there is a set of dialog acts, which is defined in a domain-independent manner. Dialog act (DA) represents the communicative function of an utterance \cite{stolcke2000dialogue} and provides information for understanding dialog content and generating responses. In general, there are three stages involved in performing DA recognition from speech: i) speech recognition, ii) utterance segmentation into DA units, and iii) DA classification on each segment. Speech recognition has been a bottleneck in the overall performance, but drastically improved by deep-learning-based approaches. End-to-end acoustic-to-word models \cite{audhkhasi2017direct,ueno2018acoustic} have shown competitive performance on large-vocabulary speech recognition tasks with an extremely simple architecture and provide a word representation for subsequent tasks. The end-to-end model can directly learn the mapping from speech to transcript without an external language model and dictionary. On the other hand, DA segmentation can be formulated as a sequence labeling task, and DA classification can be formulated as a sequence classification task. In this paper, we refer to the process of DA segmentation and classification as DA recognition as they are often jointly formulated. The DA recognition can be efficiently formulated using LSTM-based neural networks, but it is usually separately trained using a ground-truth transcript and applied to the ASR transcript.
  
  In this pipeline framework, errors from the upstream ASR still significantly degrade the performance of the whole system. Moreover, acoustic input is not considered in the subsequent component. In this study, we design a unified framework that can perform SLU along with ASR to mitigate the problems of the conventional pipeline approach.
  
  To deal with ASR errors, we integrate the ASR model with DA recognition models into a single network. The vector representing each word derived by acoustic-to-word ASR (ASR feature) is directly used in the downstream module without passing an erroneous ASR result. Moreover, acoustic input contains important information, including pause duration, for DA segmentation. We also propose a direct incorporation of DA segmentation into speech recognition by adding a segment ending label. The acoustic-to-word model performs segmentation while inferring text, utilizing both lexical and acoustic features. Finally, a full automatic system is designed and implemented for speech recognition, DA segmentation and classification. The unified model is evaluated on the Switchboard corpus.
  
  This paper is organized as follows. Section 2 explains baseline models used in our methods and experiments. Section 3 and 4 describes our proposed model to jointly predict DA segments and tags from speech. Section 5 presents experimental results of the unified models and comparison with baseline results. Section 6 discusses some related works on DA segmentation and classification from text and speech.
  
  \section{Baseline models}
  
  In this section, we introduce an approach from recent works for ASR, DA segmentation and classification tasks.
  
  \subsection{Attention-based speech recognition model}
  
  ASR with attention-based encoder-decoder model, which directly maps acoustic features into a character or word sequence without a pronunciation lexicon and a language model, has recently been intensively investigated \cite{chorowski2014end,chan2016listen,bahdanau2016end}. Besides having strong language modeling capacity, acoustic-to-word attention-based model also allows for direct connection to subsequent NLU tasks.
  
  Let $\pmb{X} = (x_1, x_2, \dots, x_T)$ denote a length-$T$ sequence of input acoustic features and $\pmb{W} = (w_1, w_2, \dots, w_L)$ denote a length-$L$ sequence of target words. The attention-based encoder-decoder model consists of two distinct subnetworks: an encoder which transforms the acoustic features into an intermediate representation $\pmb{H}=(h_1, h_2, \dots, h_{T})$ and a decoder which infers a label sequence. At the $l$-th decoding time step, a hidden state of the decoder $s_l$ is computed via a recurrent LSTM,
  \begin{equation}s_l = \text{Recurrency}(s_{l-1}, c_l, w_{l-1})\label{eq:regulardecoder}\end{equation}
  where $c_l$ denotes the context vector at the $l$-th time step and $w_{l-1}$ denotes the embedding of the last predicted label at the previous step. To compute the context vector, we use the location-aware attention introduced by \cite{chorowski2015attention}. The next output label is sampled as follows:
  \begin{equation}o_l=\text{Output}(s_l), w_l\sim\text{Generate}(o_l)\end{equation}
  where $o_l$ is an output of the decoder LSTM. A fully-connected layer is used to generate probability distribution for all words $w_l$ from the output of the decoder.
  
  \subsection{Dialog act (DA) classification model}\label{da_model}
  
  DA classification is performed on an utterance or a word sequence $W=\{w_1, w_2,...,w_L\}$. We use the term 'DA segment' and 'utterance' interchangeably and refer to a continuous sequence of utterances by a speaker as 'turn'.
  
  In the DA classification model, each word $w_{l}$ is embedded and fed into a word-level encoder, which updates its latent state $\sigma_l$ via a recurrent LSTM.
  \begin{equation}\label{eqn:da_class}\sigma_{l}=\text{Recurrency}(\sigma_{l - 1}, w_{l})\end{equation}
  Let $\sigma^k_{L_k}$ be the final decoder state of the $k$-th utterance. Each $\sigma^k_{L_k}$ can be used as an embedding vector for an utterance. By inputting this embedding to an RNN-based utterance-level encoder, we can obtain a vector $\Sigma_k$ representing the most recent $h$ segments, where $h$ is the length of history ending at the current utterance.
  \begin{equation}\Sigma_k=\text{Recurrency}(\Sigma_{k-1}, \sigma^k_{L_k})\end{equation}
  This vector $\Sigma_k$ is used to derive the DA tag $y_{DA}$ with a fully-connected layer.
  
  
  
  
  \section{Unified ASR and DA classification model}
  
  In this section, we present a unified architecture for ASR and DA classification given DA segments.
  
  \subsection{Unified ASR and DA classification by using ASR feature}
  
  Instead of using word embedding of the ASR hypothesis as input for DA classification model, we use the decoder LSTM output of the acoustic-to-word ASR model $o_l$, which we call ASR feature. In the ASR decoder, this feature is fed into a fully-connected layer to produce the probability of each label for the next recognized word. The encoder state of DA classifier is now formulated as follow:
  \begin{equation}\sigma_l=\text{Recurrency}(\sigma_{l-1}, o_l)\label{eqn:asrfeature}\end{equation}
  Here, $o_l$ is used instead of $w_l$ in equation (\ref{eqn:da_class}), which is available from the ASR internal state. The architecture of this model is shown in Fig. \ref{fig:combine} (without bold arrows).
  This ASR feature is expected to be robust against ASR errors when directly used with the DA classifier. Even if the recognized word is incorrect, the ASR feature is still expected to have a close representation to the feature of a correct word. In the conventional scheme using word embedding of recognized words, errors in ASR transcript may mislead the DA classifier into incorrect tags.
  
  This coupling can be done step by step. First, the acoustic-to-word ASR model is trained. Then, the DA classification model is trained with the ASR feature derived from the latent state of the ASR decoder. Note that this step-wise training is possible by using acoustic-to-word ASR, instead of grapheme or character-based ASR models
  
  
  
  %
  
  \subsection{End-to-end training}
  
  Furthermore, we can explore an end-to-end fine-tuning of the entire network. It adopts multi-task training that optimizes the following loss:
  \begin{equation}L=\lambda L_{DA}+(1-\lambda) L_{ASR}\end{equation}
  $L_{ASR}$ and $L_{DA}$ are cross entropy losses between the one-hot ground-truth labels and output probability distributions of ASR and DA model, respectively. 
  As degradation in ASR results can significantly impact the results of DA classifier, a small $\lambda$ is preferred.
  
  
  
  \subsection{Hybrid with ASR transcript word embedding}
  \begin{figure}[t]
    \centering
    \centerline{\includegraphics[width=0.5\textwidth]{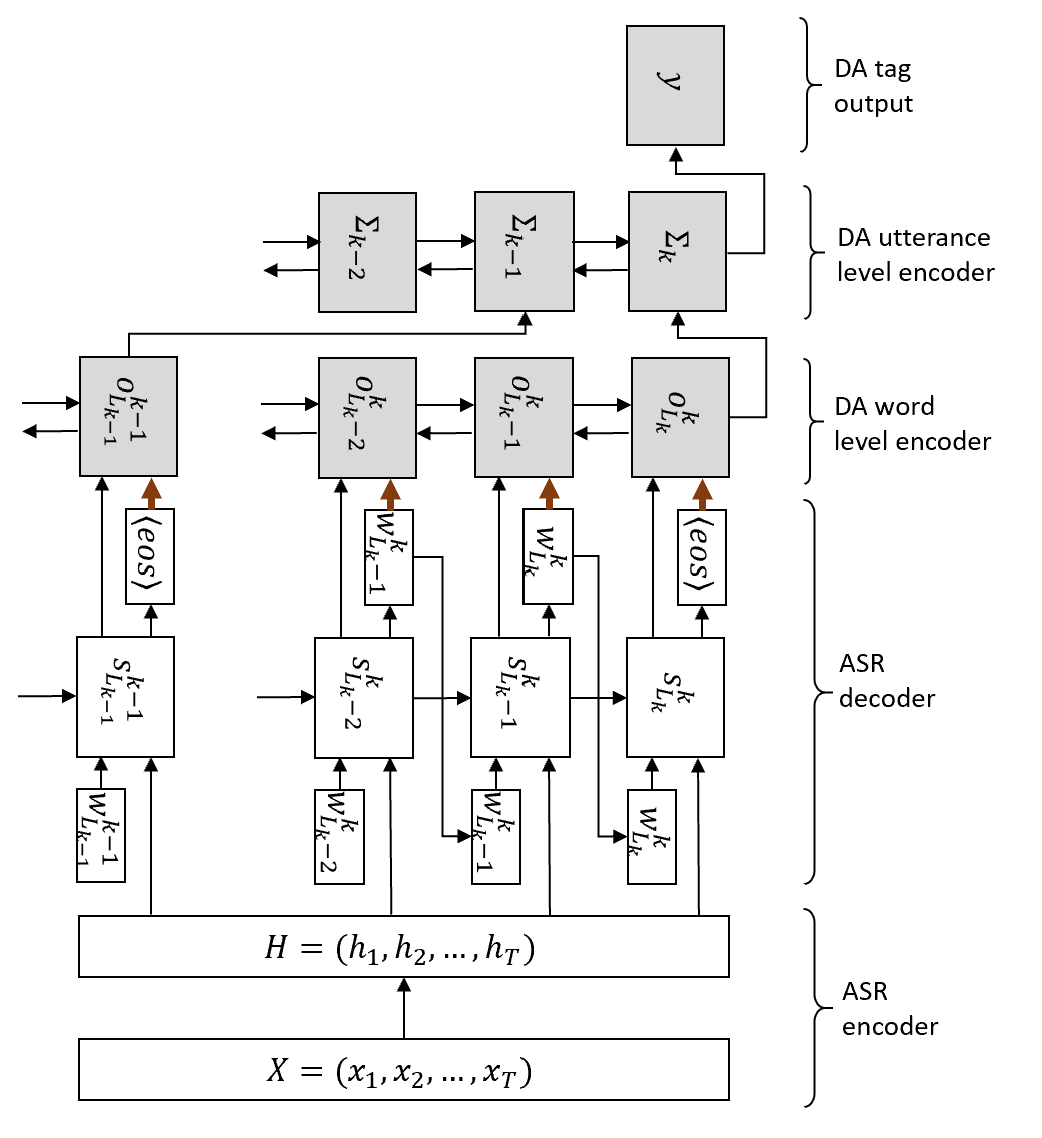}}
  \caption{Unified ASR and DA classification model. ASR feature is used as input for the DA classification model. Regular embedding (bold arrow) can also be combined in the hybrid model.}
  \label{fig:combine}
  \end{figure}  
  In the proposed model, a distributed representation of the ASR feature is expected to be robust against ASR errors, but it may not provide effective discriminant information compared with the conventional model trained with the ground-truth transcript. When there are few errors in the ASR hypothesis, a regular word embedding will give a better representation of words of an utterance. Therefore, a combined vector of the regular word embedding and the ASR feature is proposed to take advantage of both representations. The formula (\ref{eqn:asrfeature}) becomes
  \begin{equation}\sigma_l=\text{Recurrency}(\sigma_{l-1}, o_l+w_l)\end{equation}
  where $w_l$ is the word embedding learned from the ground-truth corpus by the DA classification model. The bold arrows in Fig. \ref{fig:combine} indicate this combination.
  
  
  %
  
  \subsection{Joint model with DA segmentation}
  
  In this section, we extend the framework to joint DA segmentation and classification. We introduce a DA boundary symbol to conduct DA segmentation during the ASR inference phase. 
  A DA ending symbol is added into the vocabulary and inserted after every utterance in the training dataset. 
  The ASR model transcribes speech to text and performs DA segmentation in a single step. This incorporation also allows DA segmentation to be learned from the acoustic inputs. 
  A DA classifier is coupled with the ASR model. 
  The ASR feature obtained from the acoustic-to-word ASR model is further fed into the utterance-level encoder to derive a DA tag at each DA ending position. 
  Hereafter, we use "unified model" to refer to this final model.

  \section{Experimental evaluations}
  
  \subsection{Dataset}
  
  A dataset with DA annotation and speech transcript is required for the experiments. We use the Switchboard corpus. The Switchboard Dialog Act corpus (SwDA) is annotated based on Penn Treebank 3 parses for a part of the Switchboard corpus. The two resources are aligned to obtain utterances with both transcript in the spoken domain (without punctuation or social signals) and DA tags. The combined dataset comprises 1,126 conversations with 213.5k utterances and the vocabulary size of 27.2k. There are 43 DA tags in total. We chose 20 conversations for the test set and 40 conversations for the validation set. Table \ref{tab:split} presents details on the dataset.

  \begin{table}[htb]
  \caption{Dataset overview}\label{tab:split}
  \centering
  \begin{tabular}{lccc}
  \toprule
   & train & validation & test \\ \midrule
  \#conversations    & 1066  & 40        & 20 \\
  \#DA segments    & 200.2k  & 8.9k        & 4.4k \\ 
  \#words & 1350.7k & 57.3k & 28.9k \\ \bottomrule
  \end{tabular}
  \end{table}
  
  \subsection{System configurations}
  
  A 120-dimensional feature vector of 40-channel log Mel-scale filterbank (lmfb) outputs and their delta and acceleration coefficients are
  used as acoustic features. In the ASR model, the acoustic encoder consists of 3 layers of bidirectional LSTM with 512 cells. 
  The word decoder consists of a single-layer LSTM with 512 cells. 
  ASR is conducted by simple 1-best search and also $n$-best beam search.
  
  In the DA classification model, the word-level encoder consists of 2 layers of bi-directional LSTM with 128 cells. The utterance-level encoder is a single-layer LSTM with 256 cells. A fully-connected layer is used to infer a DA class from the utterance-level encoder's output. ASR $n$-best hypotheses are combined to infer the DA tag when using beam search. Note that the ASR feature is also computed for each hypothesis of the context. 
  
  We use word error rate (WER) to evaluate ASR performance and label error rate (LER) to evaluate the results of DA classification. We pretrain an acoustic-to-word attention-based model with the training set. WER is 27.76\% with a beam search decoder (beam width 5). This model is used as the baseline and to extract the ASR feature for following experiments. 
  On the speech transcript, the DA classification model achieves an LER of 26.79\% and 25.06\% with history length of 1 and 5 ($h=1$ and $5$). 
  When evaluated on the ASR transcript with 27.76\% WER, these error rates increase to 31.85\% and 31.10\%. Models are implemented using Tensorflow\footnote{The code is published on Github \url{https://github.com/Kyoto-University-Speech-and-Audio/speech-to-dialog-act}}.
  
  
  
  \begin{table*}[t]
  \caption{DA classification error rates of baseline and unified model with given segmentation.}
  \medskip\centering
  \begin{tabular}{lcccccc}
  \toprule
  Model & \multicolumn{3}{c}{1-best search} & 
  \multicolumn{3}{c}{$n$-best beam search}\\ \midrule
  
  & WER &  {\renewcommand{\arraystretch}{0.8}\begin{tabular}[x]{@{}c@{}}LER\\\tiny{($h=1$)}\end{tabular}} & {\renewcommand{\arraystretch}{0.8}\begin{tabular}[x]{@{}c@{}}LER\\\tiny{($h = 5$)}\end{tabular}}
  & WER & {\renewcommand{\arraystretch}{0.8}\begin{tabular}[x]{@{}c@{}}LER\\\tiny{($h=1$)}\end{tabular}} & {\renewcommand{\arraystretch}{0.8}\begin{tabular}[x]{@{}c@{}}LER\\\tiny{($h = 5$)}\end{tabular}} \\ \midrule
  
  Baseline (ground-truth text) & 0.00 & 26.79 & 25.06 
  & 0.00 & 26.79 & 25.06 \\
  Baseline (ASR transcript) & 31.04 & 33.62 & 31.62 
  & 27.76 & 31.85 & 31.10 \\ \midrule
  Training with ASR transcript & 31.04 & 32.87 & 32.42 
  & 27.76 & 32.03 & 30.67 \\
  Training with ASR feature & 31.04 & 32.83 & 31.24 
  & 27.76 & 30.92 & 28.81 \\ 
  + Joint training ($\lambda=0.1$) & 35.52 / 34.74 & \bf{31.92} & \bf{29.60} 
  & 30.41 / 30.39 & 30.94 & 28.90 \\
  + Hybrid with ASR transcript & 31.04 & 32.78 & 30.83 
  & 27.76 & \bf{30.49} & \bf{28.69} \\  \bottomrule
  \end{tabular}
  \label{tab:cls_result}
  \end{table*}
  
  
  
  \subsection{Evaluation of DA classification model}
  
  Table \ref{tab:cls_result} shows the result of the baseline and the unified DA classification model with different settings. When trained with the ASR transcript, although the DA classifier is expected to adapt to the ASR output, it did not improve the accuracy in every case. A larger improvement is obtained by the proposed unified model using the ASR feature, which reduces the DA classification error rate. 
  Note that the use of beam search and $n$-best hypotheses is more effective with the ASR feature. This is probably because the improved ASR performance with the beam search provides a better ASR feature representation. The improvement is statistically significant with $h=5$.
  
  The joint training with the ASR model gives the best results with the 1-best search, which are statistically significant compared to training with ASR features ($h=5$) or baseline results. When using beam search and $n$-best hypotheses, however, the effect of the joint training is not observed. This is because the joint training considers only the 1-best ASR hypothesis in DA recognition and the loss function. An improvement in the ASR result by the beam search has a better impact than the joint training technique. The hybrid system gives a further small accuracy improvement. This suggests using the $n$-best ASR feature is almost sufficient. In total, the final system performance (28.69\%) achieves an improvement of 2.41\% absolute (7.75\% relative) from the baseline (31.10\%).
  
  \subsection{Evaluation of joint model with DA segmentation}
  
  Next, we conduct evaluations of the joint model with DA segmentation. The baseline system has 36.28\% WER when being trained and evaluated on a turn (longer than an utterance). 
  
  Table \ref{tab:asr_seg_result} presents the results with two evaluation metrics. As DA segmentation is performed on inputs which do not necessarily have the same length as the ground-truth, we define segment error rate (SER), which is calculated as the regularized sum of distances between each DA ending position in the ASR result and the closest position in the ground-truth transcript. Let $G$ and $P$ are sets of indices for a DA ending symbol in each sentence. The distance is calculated as follow.
\begin{equation}d(G,P)=\frac{1}{2}\left(\sum_{g\in G}\min_{p\in P}|g-p|+\sum_{p\in P}\min_{g\in G}|p-g|\right)\end{equation}
With this evaluation, $d(G, P)=0$ if and only if two texts have the same length and segments. We also assess the model in terms of the number of segments (NSER). The error rate is calculated by $\frac{|N^* - N|}{N}$, where $N$ and $N^*$ are numbers of DA segments of the ground-truth and the recognized segmentation, and does not consider the position of these segments. While NSER only evaluates the number of segments, SER takes into account the relative position of the DA ending labels. The unified model outperforms the baseline method by 4.22\% (relatively 25\%) in SER.
  
  \begin{table}[th]
  \caption{Comparison of segmentation error rates.}
  \medskip\centering
  \begin{tabular}{lccc}
  \toprule
  Model      & WER & SER & NSER \\\midrule
  Baseline (ground-truth text) & 0.00 & 4.81 & 10.33 \\
  Baseline (ASR transcript) & 36.28 & 16.85 & 15.32 \\ \midrule
  Joint model & 36.40 & \bf{12.63} & \bf{14.78} \\ \bottomrule
  \end{tabular}
  \label{tab:asr_seg_result}
  \end{table}
  
  
  
  
  
  To evaluate the joint error rate for segmentation and classification on the ground-truth text and the ASR hypothesis, we define DA error rate (DAER), which is calculated by replacing all non-tag labels with the DA tag of its segment and take the regularized edit distance. 
For example, a ground-truth input ``\textit{yeah} $\langle da\_ny\rangle$ \textit{i am a student} $\langle da\_sd\rangle$", recognized as ``\textit{yeah} $\langle da\_ny\rangle$ \textit{i the student} $\langle da\_sv\rangle$" (where \textit{ny}, \textit{sd} and \textit{sv} are DA tags),
has an edit distance of $4$ (edit distance of "\textit{ny sd sd sd sd}" and "\textit{ny sv sv sv}"), giving a DAER of $0.8$. 

Table \ref{tab:cascading_result} shows evaluations of the unified DA recognition model compared with the baseline method, which conducts ASR, 
segmentation and classification in a cascaded manner. We also conducted experiments with ground-truth inputs and realized that much of the degradation comes from ASR errors rather than segmentation errors. 
  The unified model improves the DAER by relatively 2.9\% with a history length of 5.
  \begin{table}[th]
  \caption{Comparison of joint DA error rates.}
  \medskip\centering
  \begin{tabular}{lcccc}
  \toprule
  Model & 
  WER & {\renewcommand{\arraystretch}{0.8}\begin{tabular}[x]{@{}c@{}}DAER\\\tiny{($h=1$)}\end{tabular}} & {\renewcommand{\arraystretch}{0.8}\begin{tabular}[x]{@{}c@{}}DAER\\\tiny{($h=5$)}\end{tabular}} \\ \midrule
  Ground-truth text \& segment & 
  0.00 & 
  26.27 & 24.61 \\
  Ground-truth text & 
  0.00 & 
  27.86 & 26.48 \\
  Baseline (ASR transcript) & 
  36.28 & 
  37.10 & 36.16 \\ \midrule
  Unified model & 
  36.40 & \bf{36.52} & \bf{35.10} \\
  \bottomrule
  \end{tabular}
  \label{tab:cascading_result}
  \end{table}
  
  \section{Related works}
  \label{sec:related_works}
  
  A number of works have been done for DA classification \cite{stolcke2000dialogue,joty2016speech,shen2016neural} or simultaneous DA segmentation and classification \cite{ang2005automatic,quarteroni2011simultaneous,zimmermann2009joint,zhao2017joint} on text input. It has also been shown that supplementary acoustic features can improve the performance of these models \cite{shriberg1998can,ang2005automatic,arsikere2016novel}
  
  In practice, DA segmentation and classification are performed on an ASR hypothesis. In a majority of previous works, however, text transcript is presupposed for input of the model, which ignores the impact of errors propagated from the ASR system. Some researchers have investigated the impact of ASR errors on subsequent SLU tasks. Ang et al. \cite{ang2005automatic} investigated the impact of ASR errors on the joint DA segmentation and classification task. Quarteroni et al. \cite{quarteroni2011simultaneous} compared the performance of DA segmentation and classification under different simulated ASR error rates.
  
  Few efforts have been made to mitigate the impact of errors from ASR. Traditionally, use of multiple ASR hypotheses were investigated. Hakkani-T{\"u}r et al. \cite{hakkani2006beyond} proposed to use a word confusion network to take into consideration multiple ASR hypotheses with their word confidence scores. Recently, neural network-based architectures are investigated. Schumann and Angkititrakul \cite{schumann2018incorporating} proposed an attention-based error correction component to improve performance of ASR, which results in better accuracy for joint intent detection and slot filling. Recently, some end-to-end approaches are investigated to perform intent classification from speech without speech-to-text conversion \cite{chen2018slu,serdyuk2018slu,bhosale2019end}. Haghani et al. \cite{haghani2018audio} investigated several architectures for end-to-end SLU incorporated with ASR and reported some improvements by jointly-trained and multi-stage unified models. Lugosch et al. \cite{lugosch2019speech} and Caubrière et al. \cite{caubriere2019curriculum} addressed the effectiveness of pre-training and transfer learning techniques for low-resource SLU.
  
  Our work has several major differences from previous works on end-to-end speech understanding. While a majority of works deal with a limited set of intents such as simple command \& control tasks \cite{haghani2018audio,lugosch2019speech}, customer care call \cite{chen2018slu} or booking and reservation \cite{caubriere2019curriculum}, we address
  general dialogue act recognition using the Switchboard corpus, which
  is widely used in the community. Moreover, our model uses
  acoustic-to-word ASR instead of character-based or subword-based ASR
  to capture lexical-level information for the interface of ASR and NLU. We further propose the hybrid features which result in the best performance. Finally, we also incorporate joint segmentation into the semantic unit, which was not addressed in the previous works.
  
  \section{Conclusions}
  
  We have presented an approach to a unified ASR and DA recognition. To our best knowledge, this is the first work to propose an end-to-end model from speech to DA. The unification of two models benefits from the ASR feature, joint training and hybrid with the ASR transcript. When using the $n$-best beam search, these methods achieved an improvement of absolute 2.41\% in DA classification accuracy. 

\bibliographystyle{IEEEtran}

\bibliography{mybib}

\begin{thebibliography}{10}
\providecommand{\url}[1]{#1}
\csname url@samestyle\endcsname
\providecommand{\newblock}{\relax}
\providecommand{\bibinfo}[2]{#2}
\providecommand{\BIBentrySTDinterwordspacing}{\spaceskip=0pt\relax}
\providecommand{\BIBentryALTinterwordstretchfactor}{4}
\providecommand{\BIBentryALTinterwordspacing}{\spaceskip=\fontdimen2\font plus
\BIBentryALTinterwordstretchfactor\fontdimen3\font minus
  \fontdimen4\font\relax}
\providecommand{\BIBforeignlanguage}[2]{{%
\expandafter\ifx\csname l@#1\endcsname\relax
\typeout{** WARNING: IEEEtran.bst: No hyphenation pattern has been}%
\typeout{** loaded for the language `#1'. Using the pattern for}%
\typeout{** the default language instead.}%
\else
\language=\csname l@#1\endcsname
\fi
#2}}
\providecommand{\BIBdecl}{\relax}
\BIBdecl

\bibitem{stolcke2000dialogue}
A.~Stolcke, K.~Ries, N.~Coccaro, E.~Shriberg, R.~Bates, D.~Jurafsky, P.~Taylor,
  R.~Martin, C.~V. Ess-Dykema, and M.~Meteer, ``Dialogue act modeling for
  automatic tagging and recognition of conversational speech,''
  \emph{Computational linguistics}, vol.~26, no.~3, pp. 339--373, 2000.

\bibitem{audhkhasi2017direct}
K.~Audhkhasi, B.~Ramabhadran, G.~Saon, M.~Picheny, and D.~Nahamoo, ``Direct
  acoustics-to-word models for english conversational speech recognition,''
  \emph{Proc. Interspeech 2017}, pp. 959--963, 2017.

\bibitem{ueno2018acoustic}
S.~Ueno, H.~Inaguma, M.~Mimura, and T.~Kawahara, ``Acoustic-to-word
  attention-based model complemented with character-level ctc-based model,'' in
  \emph{Proc. IEEE-ICASSP}, 2018.

\bibitem{chorowski2014end}
J.~Chorowski, D.~Bahdanau, K.~Cho, and Y.~Bengio, ``End-to-end continuous
  speech recognition using attention-based recurrent nn: First results,'' in
  \emph{NIPS 2014 Workshop on Deep Learning, December 2014}, 2014.

\bibitem{chan2016listen}
W.~Chan, N.~Jaitly, Q.~Le, and O.~Vinyals, ``Listen, attend and spell: A neural
  network for large vocabulary conversational speech recognition,'' in
  \emph{Acoustics, Speech and Signal Processing (ICASSP), 2016 IEEE
  International Conference on}.\hskip 1em plus 0.5em minus 0.4em\relax IEEE,
  2016, pp. 4960--4964.

\bibitem{bahdanau2016end}
D.~Bahdanau, J.~Chorowski, D.~Serdyuk, P.~Brakel, and Y.~Bengio, ``End-to-end
  attention-based large vocabulary speech recognition,'' in \emph{Acoustics,
  Speech and Signal Processing (ICASSP), 2016 IEEE International Conference
  on}.\hskip 1em plus 0.5em minus 0.4em\relax IEEE, 2016, pp. 4945--4949.

\bibitem{chorowski2015attention}
J.~K. Chorowski, D.~Bahdanau, D.~Serdyuk, K.~Cho, and Y.~Bengio,
  ``Attention-based models for speech recognition,'' in \emph{Advances in
  neural information processing systems}, 2015, pp. 577--585.

\bibitem{joty2016speech}
S.~Joty and E.~Hoque, ``Speech act modeling of written asynchronous
  conversations with task-specific embeddings and conditional structured
  models,'' in \emph{Proceedings of the 54th Annual Meeting of the Association
  for Computational Linguistics (Volume 1: Long Papers)}, vol.~1, 2016, pp.
  1746--1756.

\bibitem{shen2016neural}
S.-s. Shen and H.-Y. Lee, ``Neural attention models for sequence
  classification: Analysis and application to key term extraction and dialogue
  act detection,'' \emph{Interspeech 2016}, pp. 2716--2720, 2016.

\bibitem{ang2005automatic}
J.~Ang, Y.~Liu, and E.~Shriberg, ``Automatic dialog act segmentation and
  classification in multiparty meetings,'' in \emph{Acoustics, Speech, and
  Signal Processing, 2005. Proceedings.(ICASSP'05). IEEE International
  Conference on}, vol.~1.\hskip 1em plus 0.5em minus 0.4em\relax IEEE, 2005,
  pp. I--1061.

\bibitem{quarteroni2011simultaneous}
S.~Quarteroni, A.~V. Ivanov, and G.~Riccardi, ``Simultaneous dialog act
  segmentation and classification from human-human spoken conversations,'' in
  \emph{Acoustics, Speech and Signal Processing (ICASSP), 2011 IEEE
  International Conference on}.\hskip 1em plus 0.5em minus 0.4em\relax IEEE,
  2011, pp. 5596--5599.

\bibitem{zimmermann2009joint}
M.~Zimmermann, ``Joint segmentation and classification of dialog acts using
  conditional random fields,'' in \emph{Tenth Annual Conference of the
  International Speech Communication Association}, 2009.

\bibitem{zhao2017joint}
T.~Zhao and T.~Kawahara, ``Joint learning of dialog act segmentation and
  recognition in spoken dialog using neural networks,'' in \emph{Proceedings of
  the Eighth International Joint Conference on Natural Language Processing
  (Volume 1: Long Papers)}, vol.~1, 2017, pp. 704--712.

\bibitem{shriberg1998can}
E.~Shriberg, A.~Stolcke, D.~Jurafsky, N.~Coccaro, M.~Meteer, R.~Bates,
  P.~Taylor, K.~Ries, R.~Martin, and C.~Van Ess-Dykema, ``Can prosody aid the
  automatic classification of dialog acts in conversational speech?''
  \emph{Language and speech}, vol.~41, no. 3-4, pp. 443--492, 1998.

\bibitem{arsikere2016novel}
H.~Arsikere, A.~Sen, A.~Prathosh, and V.~Tyagi, ``Novel acoustic features for
  automatic dialog-act tagging,'' in \emph{Acoustics, Speech and Signal
  Processing (ICASSP), 2016 IEEE International Conference on}.\hskip 1em plus
  0.5em minus 0.4em\relax IEEE, 2016, pp. 6105--6109.

\bibitem{hakkani2006beyond}
D.~Hakkani-T{\"u}r, F.~B{\'e}chet, G.~Riccardi, and G.~Tur, ``Beyond asr
  1-best: Using word confusion networks in spoken language understanding,''
  \emph{Computer Speech \& Language}, vol.~20, no.~4, pp. 495--514, 2006.

\bibitem{schumann2018incorporating}
R.~Schumann and P.~Angkititrakul, ``Incorporating asr errors with
  attention-based, jointly trained rnn for intent detection and slot filling,''
  in \emph{2018 IEEE International Conference on Acoustics, Speech and Signal
  Processing (ICASSP)}.\hskip 1em plus 0.5em minus 0.4em\relax IEEE, 2018, pp.
  6059--6063.

\bibitem{chen2018slu}
Y.~Chen, R.~Price, and S.~Bangalore, ``Spoken language understanding without
  speech recognition,'' in \emph{2018 IEEE International Conference on
  Acoustics, Speech and Signal Processing (ICASSP)}.\hskip 1em plus 0.5em minus
  0.4em\relax IEEE, 2018, pp. 6189--6193.

\bibitem{serdyuk2018slu}
D.~Serdyuk, Y.~Wang, C.~Fuegen, A.~Kumar, B.~Liu, and Y.~Bengio, ``Towards
  end-to-end spoken language understanding,'' in \emph{2018 IEEE International
  Conference on Acoustics, Speech and Signal Processing (ICASSP)}.\hskip 1em
  plus 0.5em minus 0.4em\relax IEEE, 2018, pp. 5754--5758.

\bibitem{bhosale2019end}
S.~Bhosale, I.~Sheikh, S.~H. Dumpala, and S.~K. Kopparapu, ``End-to-end spoken
  language understanding: Bootstrapping in low resource scenarios,''
  \emph{Proc. Interspeech 2019}, pp. 1188--1192, 2019.

\bibitem{haghani2018audio}
P.~Haghani, A.~Narayanan, M.~Bacchiani, G.~Chuang, N.~Gaur, P.~Moreno,
  R.~Prabhavalkar, Z.~Qu, and A.~Waters, ``From audio to semantics: Approaches
  to end-to-end spoken language understanding,'' in \emph{2018 IEEE Spoken
  Language Technology Workshop (SLT)}.\hskip 1em plus 0.5em minus 0.4em\relax
  IEEE, 2018, pp. 720--726.

\bibitem{lugosch2019speech}
L.~Lugosch, M.~Ravanelli, P.~Ignoto, V.~S. Tomar, and Y.~Bengio, ``Speech model
  pre-training for end-to-end spoken language understanding,'' \emph{arXiv
  preprint arXiv:1904.03670}, 2019.

\bibitem{caubriere2019curriculum}
A.~Caubri{\`e}re, N.~Tomashenko, A.~Laurent, E.~Morin, N.~Camelin, and
  Y.~Est{\`e}ve, ``Curriculum-based transfer learning for an effective
  end-to-end spoken language understanding and domain portability,''
  \emph{arXiv preprint arXiv:1906.07601}, 2019.

\end{thebibliography}


\end{document}